\documentclass[prl,twocolumn,superscriptaddress,floatfix,noshowpacs,10pt]{revtex4-1}%
\usepackage{graphicx,bm,times}
\usepackage{amsmath}
\usepackage{amsfonts}
\usepackage{amssymb}

\begin{document}

\title{Emergence of the nematic electronic state in FeSe}

\author{M. D. Watson}
\affiliation{Clarendon Laboratory, Department of Physics,
University of Oxford, Parks Road, Oxford OX1 3PU, UK}

\author{T. K. Kim}
\affiliation{Diamond Light Source, Harwell Campus, Didcot, OX11 0DE, UK}

\author{A. A. Haghighirad}
\affiliation{Clarendon Laboratory, Department of Physics,
University of Oxford, Parks Road, Oxford OX1 3PU, UK}

\author{N. R. Davies}
\affiliation{Clarendon Laboratory, Department of Physics,
University of Oxford, Parks Road, Oxford OX1 3PU, UK}

\author{A. McCollam}
\affiliation{High Field Magnet Laboratory, Institute
for Molecules and Materials, Radboud University, 6525 ED Nijmegen, The Netherlands}

\author{A. Narayanan}
\affiliation{Clarendon Laboratory, Department of Physics,
University of Oxford, Parks Road, Oxford OX1 3PU, UK}

\author{S. F. Blake}
\affiliation{Clarendon Laboratory, Department of Physics,
University of Oxford, Parks Road, Oxford OX1 3PU, UK}

\author{Y.~L.~Chen}
\affiliation{Clarendon Laboratory, Department of Physics,
University of Oxford, Parks Road, Oxford OX1 3PU, UK}

\author{S. Ghannadzadeh}
\affiliation{High Field Magnet Laboratory, Institute
for Molecules and Materials, Radboud University, 6525 ED Nijmegen, The Netherlands}

\author{A.~J.~Schofield}
\affiliation{School of Physics and Astronomy, University of Birmingham,
Edgbaston, Birmingham B15 2TT, UK}

\author{M. Hoesch}
\affiliation{Diamond Light Source, Harwell Campus, Didcot, OX11 0DE, UK}

\author{C. Meingast}
\affiliation{Institute for Solid State Physics, Karlsruhe Institute of Technology, Germany}

\author{T. Wolf}
\affiliation{Institute for Solid State Physics, Karlsruhe Institute of Technology, Germany}

\author{A. I. Coldea}
\email[corresponding author:]{amalia.coldea@physics.ox.ac.uk}
\affiliation{Clarendon Laboratory, Department of Physics,
University of Oxford, Parks Road, Oxford OX1 3PU, UK}

\begin{abstract}
We present a comprehensive study of the evolution of the nematic electronic structure of FeSe
using high resolution angle-resolved photoemission spectroscopy (ARPES), quantum oscillations in the normal
state and elastoresistance measurements. Our high resolution ARPES allows us to track
the Fermi surface deformation from four-fold to two-fold symmetry across the structural
transition at $\sim 87$~K %from tetragonal to orthorhombic
which is stabilized as a result of the dramatic splitting of bands associated with $d_{xz}$ and $d_{yz}$ character
in the presence of strong electronic interactions.
The low temperature Fermi surface is that of a compensated metal consisting of one hole
and two electron bands and is fully determined by combining the knowledge from ARPES and quantum oscillations.
A manifestation of the nematic state is the significant increase in the nematic susceptibility approaching the structural transition that we detect from our elastoresistance measurements on FeSe.
The dramatic changes in electronic structure cannot be explained by the small lattice
effects and, in the absence of magnetic fluctuations above the structural transition,
point clearly towards an electronically driven transition in FeSe,
stabilized by orbital-charge ordering.

\end{abstract}
\date{\today}
\maketitle
%%%%%%%%%%%%%%%%%%%%%%%%%%%%%%%%%%%%%%%%%%%%%%%%

A nematic state is a form of electronic order
which breaks the rotational symmetries without changing the translational symmetry
of the lattice, and this state may play an important role in understanding high temperature superconductivity.
There has been great interest in determining
 such electron  nematic states~\cite{fradkin_2010a}, most recently
in iron-based superconductors \cite{Fernandes2014}.
Several scenarios have been proposed to explain the observed nematicity
based on phonon-driven and electronically-driven tetragonal symmetry breaking.
As the lattice effects are very small and such a state is found in close proximity to a magnetic state,
it has been suggested that the nematicity is generated by spin fluctuations and is a precursor of the
incipient antiferromagnetic state \cite{Fernandes2014}.
Other models emphasizes the tendency towards orbital order
for the Fe $d$ electrons in different local environments,
resulting in the development of unequal occupancy of $d_{xz}$ and $d_{yz}$ orbitals which
drives the symmetry-breaking, and the magnetism is stabilized as a consequence \cite{Phillips2009}.
Electronic nematic order can be revealed through
its strongly anisotropic resistivity
and findings so far in iron-based superconductors
have supported the spin-fluctuations scenario \cite{Chu2012,Kuo2013}.
Understanding the dominant electronic interactions is of great importance,
for they could determine the symmetry and properties of the superconducting state \cite{Fernandes2014}.

With a superconducting transition of $T_c \sim 9$~K, FeSe is a special case amongst Fe-based superconductors, since it undergoes a structural transition at $T_s \sim{}87$~K but does not order magnetically 
at any temperature.  It has also attracted a lot of interest due to
the strong increase in $T_c$ to 37~K under pressure \cite{Medvedev2009}, the existence of high-$T_c$ intercalates of FeSe \cite{Clarke2013} and, that a monolayer of FeSe grown on SrTiO$_3$ could have
its superconducting transition temperature in excess of 100~K \cite{Ge2014}. The availability of high quality bulk crystals, grown using chemical vapour transport \cite{Bohmer2013}, have recently re-opened investigations into the electronic properties of FeSe. ARPES has found evidence of a large band splitting caused by orbital ordering below the structural transition \cite{Nakayama2014,Shimojima2014} but the resolution of the available data cannot clarify the changes that occur at the Fermi level. Moreover, quantum oscillation experiments at low temperatures
have detected an unusually small Fermi surface \cite{Terashima2014,Audouard2014}.
As magnetic fluctuations are detected only below the structural transition in FeSe, it is expected that they are not
the driving force for this transition \cite{Bohmer2014,Baek2014}.
Thus, the nature of the structural transition in FeSe is rather unusual
and how the electronic structure is stabilized by breaking of the rotational symmetry
could be the key to understanding its superconductivity and how it can be further enhanced.

In this paper we explore the electronic structure of FeSe by using high-resolution ARPES, quantum oscillations and elastoresistivity measurements on the same batch of high-quality FeSe single crystals. These techniques provide a comprehensive picture of the evolution of the electronic structure of FeSe from the high-temperature tetragonal phase through the fourfold-symmetry-breaking structural transition at $T_s$ and into the electronic nematic phase. We observe
strong in-plane $d$-wave like deformation of the Fermi surface from four-fold to two-fold symmetry
as a result of band splitting with $d_{xz}$ and $d_{yz}$ character.
A direct manifestation of a nematic Fermi surface of FeSe,
is the significant increase in the nematic susceptibility measured by elastoresistance measurements
when approaching $T_s$, not emerging  due to anisotropic magnetic fluctuations
but being driven by the orbital/charge degrees of freedom.
The low temperature Fermi surface, based
on our high resolution data from ARPES and quantum oscillations,
consist of an in-plane distorted quasi-two dimensional hole band and
two electron bands; the inner electron pocket is extremely elongated and quasi-two dimensional, whereas
the outer electron band, with predominantly $d_{xy}$ character, is not detected by ARPES
but is present in quantum oscillations.
The high-temperature bands have orbital dependent band renormalisation, with respect to band structure calculations, due to many-body interactions, which are particularly significant for the $d_{xy}$ band; this leads to
 a significant shrinkage of the Fermi surfaces (a factor of $\sim$5, compared with calculations). Our
 measurements also detect the band splitting induced by spin-orbit coupling ($\sim{}20$ meV at the $\Gamma$ point).

{\bf Experimental details}
Samples were grown by the KCl/AlCl$_3$ chemical vapour transport method. Magnetotransport measurements  were performed as function of in-situ rotation in high magnetic field up to 33~T at HFML in Nijmegen,
using an excitation current of 0.8~mA. Good electrical contacts were achieved by using In solder. Quantum
oscillations were observed in more than three samples with good agreement.
ARPES  measurements  were  performed  at  the  I05  beamline  of  Diamond
Light Source, UK. Single crystal samples were cleaved in-situ in a vacuum
lower than 2$\times$10$^{-10}$ mbar and measured at temperatures ranging from 6-120~K.
  Measurements  were  performed  using  linearly-polarised  synchrotron
light from 20-120 eV and employing Scienta R4000 hemispherical electron
energy  analyser  with  an  angular  resolution  of  0.2-0.5  deg  and  an  energy
resolution of ~3 meV.
Elastoresistance measurements were performed by measuring
the in-plane anisotropic transport properties while straining the crystals along
the [110] direction in the tetragonal phase, by applying 
a voltage to a piezoelectric stack, similar to those discussed in Ref. \cite{Kuo2013}
and detailed in SM.
Band  structure  calculations  were  performed  using the  experimental  crystal
structure of FeSe in Wien2k using the GGA approximation with spin-orbit coupling included.
 Since FeSe does not show magnetic order, ARPES spectra are compared to non-spin-polarized band structure,
with the relaxed lattice parameters, $a = 3.7651$ \AA, $c$ = 5.5178 \AA, $z_{Se}$ = 0.24128 \cite{Kumar2012}.

\begin{figure}[htbp]
	\centering
    \includegraphics[width=9cm]{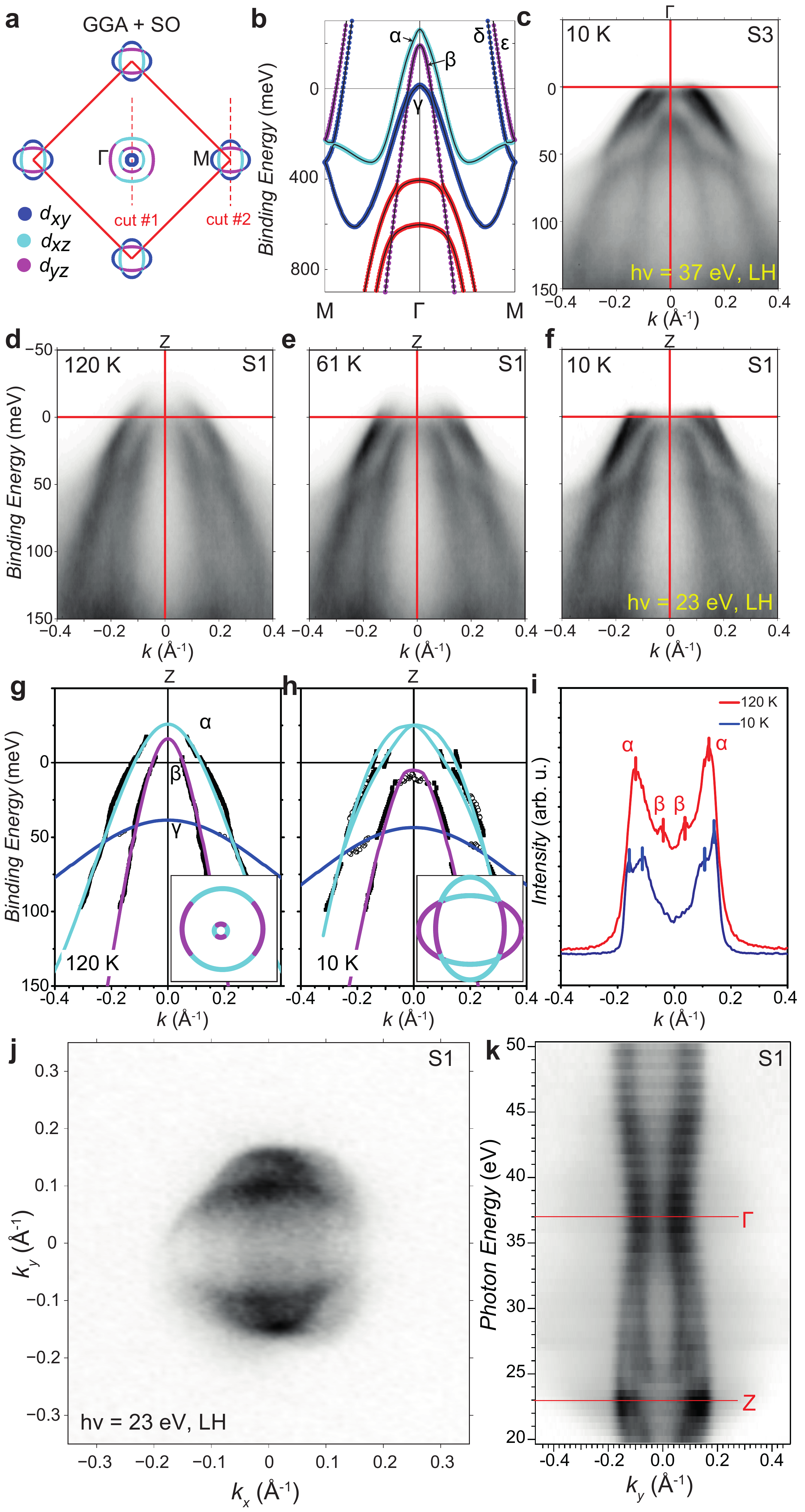}
    \caption{\textbf{Temperature dependence of the hole bands in FeSe.} a) and b) Band structure calculation of the Fermi surface of FeSe in the tetragonal phase, projected into the $k_z=0$ plane and colored by dominant orbital character; the high symmetry cuts are indicated. Calculations predict three hole pockets around the $\Gamma$ point and two electron pockets around the M point. c) The high-symmetry M-$\Gamma$-M cut (cut 1) at low temperature (10~K). d)-f) The temperature-dependence of the A-Z-A cut at the Z point ($k_z=\pi/c$ plane). g) and h) Extracted band dispersion from constrained multiple Lorentzian fits to momentum-distribution curves (MDC, solid symbols) or peaks in the energy dispersive cut (EDC, open symbols) from the data in d) and f). The solid lines in g) are renormalized band structure from a). The insets show a schematic
     Fermi surface. i) MDC for the hole bands ($\alpha$ and $\beta$) at high and low temperatures. At 10~K, the $\alpha$ splits near the Fermi level, resulting in two crossings due to formation of twin domains. j) Fermi surface map at 10~K indicating the elongation of the hole pocket combined with a duplicate rotated by 90$^\circ$ caused by twin domains. k) Photon-energy dependence of the MDC at $E_F$ (equivalent to scanning $k_z$ in the $\Gamma$-Z direction) that shows a quasi-two dimensional hole band at 10~K.}
	\label{fig1}
\end{figure}
{\bf Temperature dependence of the hole bands.}
Fig.~\ref{fig1}a shows the calculated Fermi surface of FeSe
in the tetragonal paramagnetic phase,
 which consists of three hole pockets ($\alpha$, $\beta$ and $\gamma$) around the Brillouin
 zone center ($\Gamma$) and two electron-like pockets ($\delta$ and $\epsilon$),
 similar to other reports \cite{Maletz2014}.
The bands which cross the Fermi level have mainly $d_{xz}$, $d_{yz}$ and $d_{xy}$  character (Fig.\ref{fig1}a and b).
Due to the matrix elements effects
in ARPES experiments the light polarization and the scattering geometry
allows us to select mainly either $d_{xz}$ or
$d_{yz}$ bands using LH ($p$) or LV ($s$)-polarization, respectively,
and to identify the orbital character of the measured bands, as shown in Supplementary Material (SM)
and detailed in Ref.~\onlinecite{Zhang2012b}.

Fig.~\ref{fig1}d-f show the evolution of the hole bands
by performing high-symmetry cuts, either through M-$\Gamma$-M ($k_z=0$)
or A-Z-A ($k_z=\pi/c$), as shown in Fig.~\ref{fig1}c-f as a function of temperature.
In the high-temperature tetragonal phase with preserved $C_4$ rotational symmetry,
both $\alpha$ and $\beta$ bands cross the Fermi level at the Z point, as shown by the Fermi-level momentum distribution curves (MDCs) at 120~K in Fig.\ref{fig1}i.
The degeneracy between the $\alpha$ and $\beta$ bands is lifted by the
spin-orbit coupling \cite{Cvetkovic2013} which we can directly estimate from our data
as being $\Delta_{SO} \sim 20$ ~meV, similar to results
reported for LiFeAs \cite{Borishenko2014}.
On lowering the temperature below the
structural transition at $T_s \sim 87$~K (Fig.\ref{fig1}f),
the $\beta$ band at Z is pushed below the Fermi level, as shown in Fig.~\ref{fig1}.
About $50$~meV below the Fermi level we also detect
the presence of a broad band that is
the $d_{xy}$ band, $\gamma$, in Fig.~\ref{fig1}g, having weaker intensity
 due to matrix elements and being shifted significantly
 in relation to the band structure calculations for the tetragonal structure in Fig.~\ref{fig1}b.
  The effect of the loss of $C_4$ rotational symmetry through the structural transition
  from tetragonal to orthorhombic symmetry at $T_s$
is clearly observed in the detailed Fermi surface map at the Z point, showing that the $\alpha$ pocket elongates to become elliptical, shown in Fig.\ref{fig1}h and j. Due to the natural twinning of the samples, however, we observe two ellipses (in Fig.\ref{fig1}h) rotated 90$^\circ$ with respect to each other.
The $k_z$ dependence can be determined by measuring the Fermi-level MDC
 as a function of incident photon energy, shown in Fig.~\ref{fig1}k;
we detect a quasi-two dimensional shape for the $\alpha$ hole pocket,
the only hole band that crosses the Fermi level at low temperature.
The experimental area for the hole bands, $A_k$ at 120~K is strongly reduced by a factor 5,
as compared with band structure calculations in the tetragonal phase (see SM) and
we will later discuss the consistency with our quantum
oscillations results at low temperatures (see Table~1 in SM).

\begin{figure*}[htbp]
	\centering
\includegraphics[width=17cm]{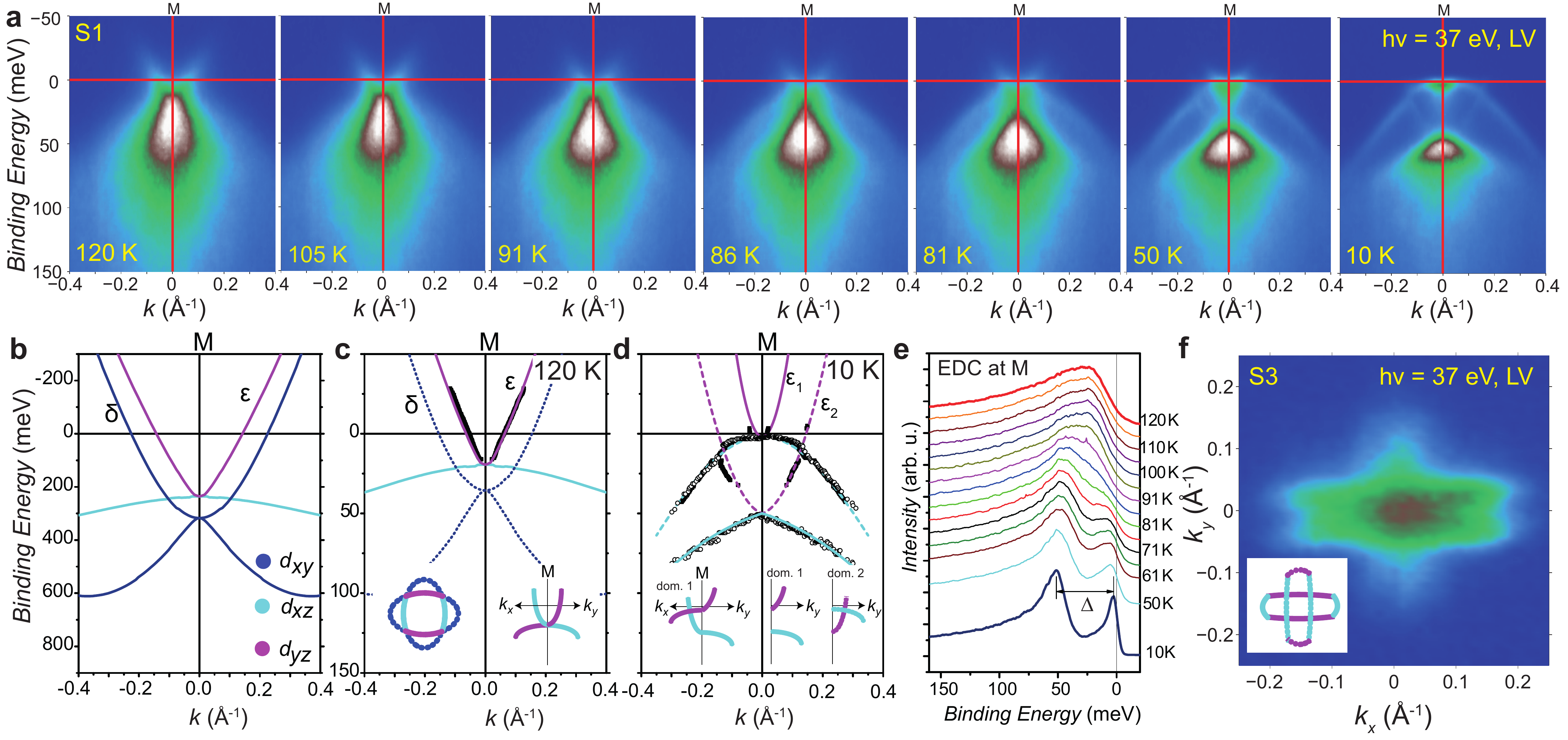}
	\caption{\textbf{Development of a strongly elongated electron Fermi surface through the structural transition.} a) Temperature dependence of high-symmetry cut through M, showing the development of a splitting of intensity which results from strong orbital-dependent shifts of the band structure with onset at $T_s$ = 87 K. b) The calculated band dispersions in the tetragonal structure. c) The experimental band structure at 120 K (solid dots) renormalized  to the high temperature data (solid lines); the outer $d_{xy}$ electron band ($\delta$) is not observed at all at the M point. d) The lifting of $d_{xz/yz}$ degeneracy at 10~K and the consequent band shifting for two different domains, $\epsilon_1$ and $\epsilon_2$, indicated by the solid and dashed lines, respectively. The insets show the schematic band dispersion for different domains and directions, as detailed in SM. e) Temperature-dependence of the EDC at the M point, showing the development of the splitting of bands with $d_{xz}$ and $d_{yz}$ character; the splitting $\Delta$ is plotted in Fig.~\ref{fig4}m. f) Fermi surface map (integrated within 2~meV of the Fermi level) showing the cross-shape arising from strongly elongated Fermi surfaces at M in the two domains, shown schematically in the inset.}
	\label{fig2}
\end{figure*}

The effect of electronic correlations  can be estimated by comparing
 band structure calculations with the band dispersion in the tetragonal phase at 120~K,
 as shown in Fig.~\ref{fig1}g.
 The obtained band renormalisation factors for FeSe are $\sim$ 3.2, 2.1 and 8 for the $\alpha$,$\beta$ and $\gamma$ pockets respectively - suggesting particularly strong orbital-dependent electronic correlations on the $\gamma$ band with $d_{xy}$ orbital character. These values are close to those reported for FeSe in Ref. \cite{Maletz2014}, whereas in the case of FeSe$_x$Te$_{1-x}$, the band-selective renormalization
 varies strongly with values between 1 and 17 \cite{Tamai2010},
the significantly higher value being assigned to the $d_{xy}$ band.
Band structure calculations including correlations (DMFT) on FeSe \cite{Yin2012}
give band renormalisations of $\sim$2.8 for the $d_{xz/yz}$ orbitals,
comparable to the measured values, whereas the
predicted value of $\sim$3.5 for the $d_{xy}$ band is smaller than in experiments.

{\bf Temperature dependence of the electron-like bands.}
We now focus on the temperature dependence of the electron-like bands in FeSe
shown in Fig.~\ref{fig2}a, visualizing the dramatic changes occurring
through the structural transition.
In the high-temperature tetragonal phase of FeSe, band structure calculations predict two cylindrical electron pockets, as shown in Fig.~\ref{fig1}a and Fig.~\ref{fig2}b.
 The outer electron-like band (labelled $\delta$) has
a dominant $d_{xy}$  character whereas the inner electron band (labelled $\epsilon$) has $d_{xz/yz}$ orbital character.
At high temperatures (120~K), the inner $d_{xz/yz}$ band has a strong intensity at the M point, where
 bands disperse both along and perpendicular to the cut direction and are four-fold degenerate,
 as shown in Fig.\ref{fig2}a.
 However, a combination of low intensity caused by matrix element effects for the
 $d_{xy}$ orbital \cite{Zhang2012b},  as well as broadening due to
  either impurity scattering \cite{Ye2014} or
 strong correlations may cause band incoherence \cite{Lanata2013,Yin2012} and difficulty to observe the $d_{xy}$ band in ARPES
at the M point \cite{Shimojima2014,Mandal2014,Tan2013}.

Following the temperature dependence of the inner electron bands with $d_{xz/yz}$ orbital character,
in the experimental $\Gamma{}$-M-$\Gamma$ cut (cut 2 in Fig.\ref{fig1}a),
the band crossing the Fermi level would have $d_{yz}$ character, whereas the downwards dispersing band
from the M point would have $d_{xz}$ character (see Fig.~\ref{fig2}c).
Below the structural transition which breaks the $C_4$ rotational symmetry,
we observe a large band splitting which can only occur if the $d_{xz/yz}$ degeneracy is broken and
 the $d_{xz}$ and $d_{yz}$ orbitals develop an unequal occupation.
As a result, the $d_{yz}$ band moves up ($\epsilon_1$ in Fig.~\ref{fig2}d),
 and the $d_{xz}$ band moves down (solid lines), but for the other structural domain the opposite occurs, the $d_{xz}$ band moves up and the $d_{yz}$ ($\epsilon_2$) moves down indicated by the dashed lines in Fig.~\ref{fig2}d and
 further detailed in SM  ($x$ and $y$ are defined in the \textit{experimental} coordinate frame).
This picture is consistent with that suggested from ARPES studies on detwinned crystals of FeSe \cite{Shimojima2014},
but the high resolution of our data allows us to detect the precise changes in the Fermi surface across the transition.
The large band splitting of  $\sim 50$~meV at the M point at 10~K
(Fig.~\ref{fig2}e) indicates the lifting of
of $d_{xz}$ and $d_{yz}$ degeneracy in FeSe. This was also observed in NaFeAs and BaFe$_2$As$_2$ over a very limited temperature range between the structural transition and magnetic transition,
\cite{Yi2011,Yi2012,He2010a,Zhang2012b}.
In FeSe, this dramatic energetic shift observed below $T_s$
 is much larger than that expected from a simple structural orthorhombic distortion of $2\times 10^{-3}$
 (($a-b$)/($a+b$)), which would cause a 17~meV shift in the absence of the renormalization effects (see SM).
 At the Fermi level, the resulting electron pocket is strongly elongated, with a $k_{F}(y)\approx 0.02 $ \AA$^{-1}$ and $k_{F}(x)\approx 0.14$ \AA$^{-1}$. However, due to twinning,  the Fermi surface consists of two ellipses at 90$^\circ$ to each other, resulting in a cross-shaped Fermi surface (Fig.~\ref{fig2}f), similar to the hole pocket.
The degree of Fermi surface distortion is measured by the temperature dependence of $k_F$,  and the ellipticity
$k_{F}(y)$($\varepsilon_1$), shown in Fig.~\ref{fig4}l, is suggestive of an order parameter of
a second order phase transition at $T_s$.
Thus, the observed experimental elongation of both the electron and hole pockets
at low temperatures in FeSe
can be thought of as a consequence of
the electronic anisotropy induced by orbital ordering
in the presence of interactions.
Another possible scenario for this behaviour
can be thought as a Pomeranchuk instability \cite{pomeranchuk_1958a}
that results in the spontaneous deformation of the Fermi surface
from a four-fold symmetric almost circular shape at high temperature to an elliptical shape at low temperatures.
In either cases, this electronically-induced Fermi surface deformation
is also supported by the small specific heat jump at $T_s$ compatible with an electronic contribution \cite{Bohmer2013}.

\begin{figure}[htbp]
	\centering
\includegraphics[width=9cm,clip=true]{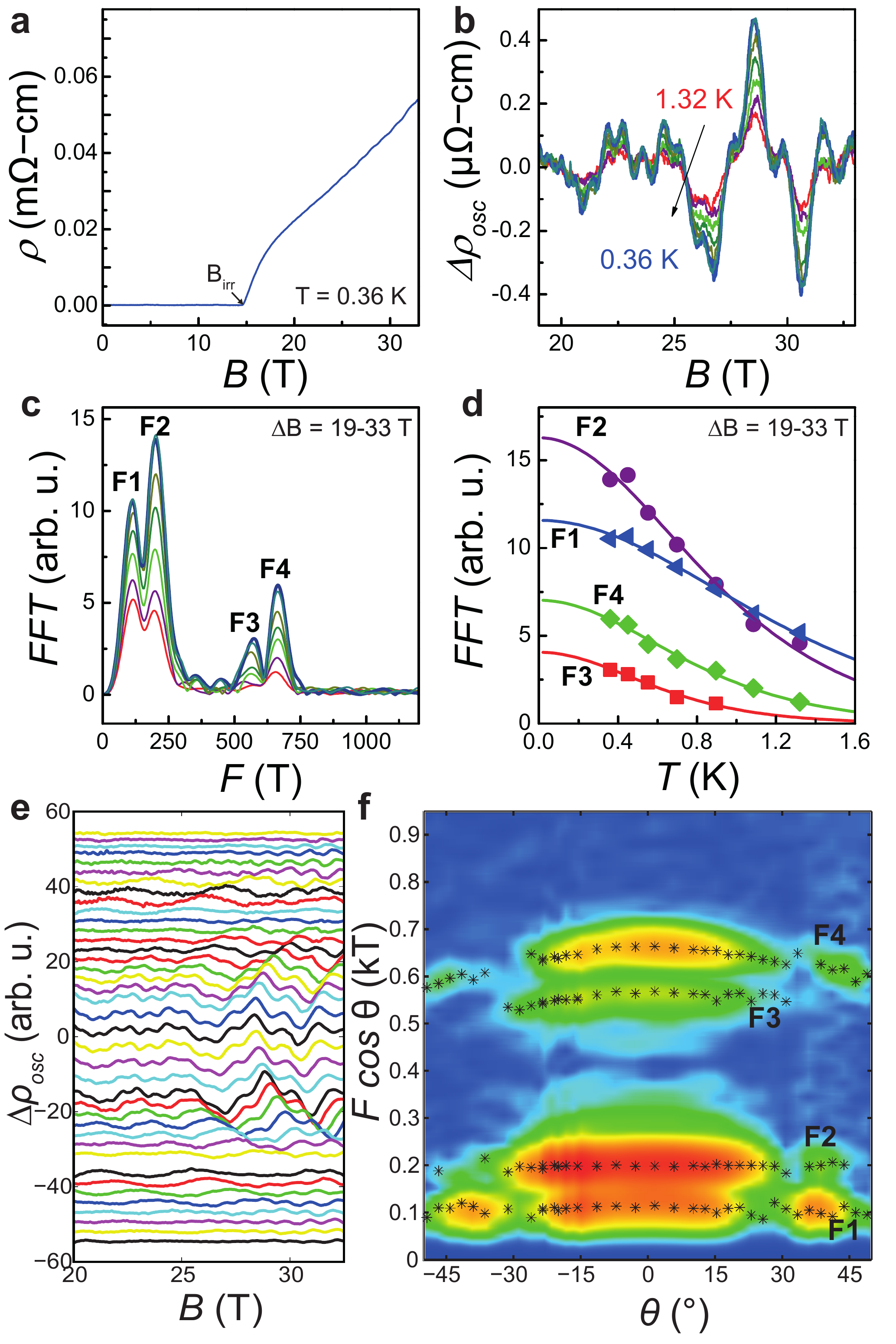}
	\caption{\textbf{Quantum oscillations in FeSe.} a) Magnetoresistance of FeSe for a field applied along the $c$ axis ($\theta$=0) with the irreversibility field, $B_{irr}$ around 14~T at 0.36~K. b) The oscillatory part of the resistivity $\Delta \rho_{osc}$ , obtained by subtracting a polynomial background from the raw data. c) Fourier transform (FFT) of the data in b) identifying four different frequencies $F_{1-4}$, corresponding to
$k$-dependent extremal areas on the Fermi surface. d) Temperature-dependence of the amplitude of oscillation, from which the effective masses may be extracted. e) Quantum oscillations and f) the corresponding Fourier transforms as a function of angle $\theta$ of applied magnetic field with respect to the $c$ axis at 0.36~K. By plotting $F \cos{\theta}$ against $\theta$ one can identify if orbits correspond to a maximum or minimum of a quasi-two dimensional Fermi surface. Color bar corresponds to the Fourier transform amplitude.}
	\label{fig3}
\end{figure}

{\bf Comparison with quantum oscillations.}
In order to provide a complete
and consistent picture of the Fermi surface of FeSe,
we have also measured quantum oscillations in
samples from the same batch used in the ARPES experiments.
Quantum oscillations are a powerful technique
that allows precise determination of cross-sectional areas
of the Fermi surface and the corresponding orbitally-averaged
quasiparticle masses but it usually needs to rely on band structure or ARPES
to provide the exact $k$-space location of these orbits.
Magnetotransport measurements are performed in the normal state of FeSe at very low temperatures and high magnetic fields, as shown in Fig.~\ref{fig3}a.
The oscillatory signal, periodic in $1/B$, is made clear by subtracting
a high order polynomial from the raw data in Fig.~\ref{fig3}b. The fast Fourier
transform of $\rho_{osc}$ shown in Fig.~\ref{fig3}c,
identifies four different quantum oscillation frequencies,
labelled $F_{1-4}$ in ascending order, that are directly linked
to the extremal areas on the Fermi surface
by the Onsager relation, $F_i = A_{ki} \hbar /(2 \pi e)$, for a particular
field orientation. All frequencies
are lower than 1~kT in agreement
with previous reports
\cite{Terashima2014,Audouard2014}.
This is smaller than the frequencies of other iron-based superconductors
that do not show any Fermi surface reconstruction, such as LaFePO \cite{Coldea2008} or LiFeAs\cite{Putzke2012}.
Band structure calculations predict the existence of five different quasi-two dimensional
cylinders with sizes (1-2.6~kT), much larger than those found in experiments (below 1~kT) (see SM).
The angular dependence of these frequencies also suggest
a quasi-two dimensional nature of these bands (see that the angular dependence of frequencies varies
almost like $1/\cos \theta $ in Fig.\ref{fig3}f), and
previous studies have suggested that at low temperatures FeSe has either
only quasi-two dimensional electron bands or electron and hole bands \cite{Terashima2014}.
\begin{figure*}[htbp]
	\centering
\includegraphics[width=0.98\linewidth]{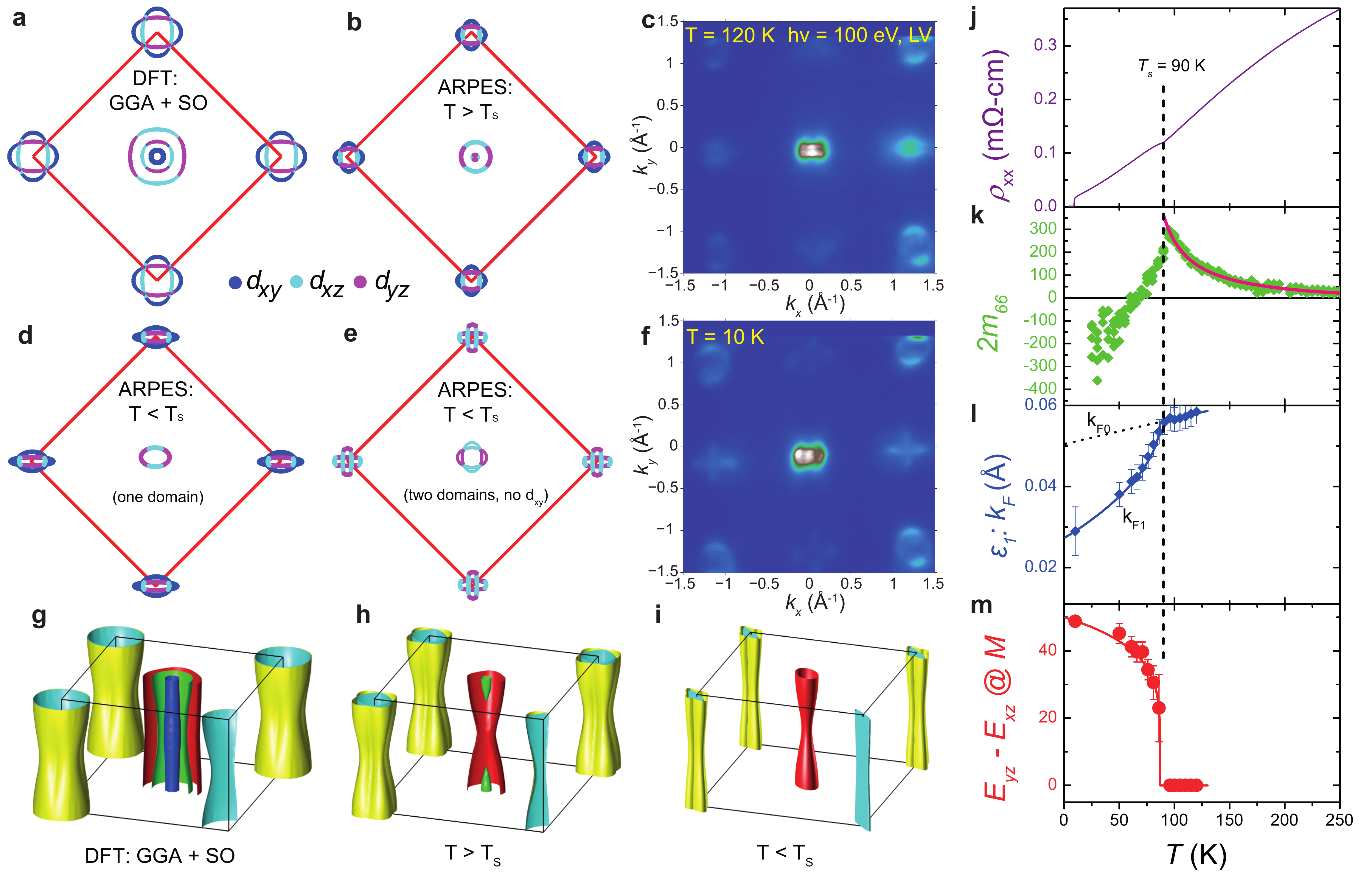}
	\caption{\textbf{Symmetry-breaking of the electronic structure of FeSe.} a) Band-structure calculation of the Fermi surface in the high-temperature $P4/nmm$ tetragonal space-group. b) Schematic and c) the experimental high-temperature in-plane Fermi surface consists of significantly shrunk pockets. d) Elongated hole and electron pockets below the structural transition. e) Schematic and f) experimental low temperature Fermi surface including the effect of twin domains (the $d_{xy}$ electron pocket is not plotted here). g-i) Three-dimensional representations of Fermi surface of FeSe, as described in a, b and d, respectively. j) Resistivity as a function of temperature for the sample used in Fig.~3. k) The induced change in resistivity to in-plane strain measured by the $m_{66}$ parameter from elastoresistance tensor in FeSe, which provides a direct measure of the electronic nematic order parameter, indicating that the structural transition is electronically-driven, as discussed in the main text. l) Temperature-dependence of the intense $k_{F1}$ for the $\epsilon$ pocket around the M point, showing the Fermi surface deformation which onsets at $T_s$. m) Energy splitting of the bands with $d_{xz}$ and $d_{yz}$ band character at the M point, extracted from Fig.~2e.}
	\label{fig4}
\end{figure*}

The cyclotron-averaged effective masses of the quasiparticles,
extracted from the temperature-dependence of the quantum oscillations
 amplitudes (Fig.\ref{fig3}d) for each orbit,
 are listed in Table~1 in SM). %\cite{Shoenberg}.
We find a reasonably good agreement between the values of the effective masses for our different samples
with those reported previously in Ref.~\onlinecite{Terashima2014,Audouard2014}.
The $F_2$ and $F_4$ orbits have similar effective masses of around 4~$m_e$, which may indicate that they originate from the same band,
 but with different $k_z$-values, as in the case of a corrugated quasi-two dimensional band, whereas
 the $F_3$ orbit has a particularly heavy effective mass of $\sim 7(1)$ $m_e$,
in agreement with other reports \cite{Terashima2014}.
However, in the case of the $F_1$
pocket there is a variation of the quasiparticle masses between
0.7(2) \cite{Audouard2014} to 3(1) $m_e$.
The strong disparity between the lighter masses $F_1$ and the heavier masses of $F_4$
points towards different band origin for these orbits at the Fermi level.

Next, we compare the absolute size of the different
$k$ values at the Fermi level extracted from ARPES with those from quantum oscillations (Table~1 in SM).
At low temperatures, ARPES detects a single quasi-two dimensional hole band (see Fig.~1k)
along the $\Gamma$(Z) direction, which has a two-fold symmetry, with a maximum area
 around Z and minimum around $\Gamma$, and a small carrier density of 2.53$\times{}10^{20}$ cm$^{-3}$.
 Similarly, ARPES gives clear indication of the presence of an electron-band centered at M
which is a strongly elongated cylinder, but it is lacking the information concerning the behaviour of the
outer electron-like $d_{xy}$ band crossing the Fermi level, which may be present according to band structure calculations.
Since ARPES data suggest that there is only one hole pocket but possibly more than one electron pocket,
 the largest maximum orbit observed in quantum oscillations, $F_4$, must corresponds to the hole pocket
 in order to maintain charge balance in the system; thus two quantum oscillation
  frequencies with similar effective masses, $F_2$ and $F_4$, belong to this quasi-two dimensional hole band, in close agreement with the size of this quasi-two dimensional band detected from ARPES (Fig.\ref{fig1}k).
Furthermore, the $F_3$ frequency in quantum oscillations has a substantially
larger effective mass than all the other frequencies, which is of similar value
to the band renormalization found in the ARPES experiments for the $d_{xy}$ band ($\gamma$ in Fig.\ref{fig1}g),
suggesting that {\it the outer electron pocket that is detected in quantum oscillations
but not in ARPES has $d_{xy}$ character}.
The inner elongated electron band from ARPES (with areas smaller than 100~T),
 can be assigned to the $F_1$ frequencies, as this band is
 almost two-dimensional (as in seen at the A-point in Fig.1 in SM).
Thus, by combining the knowledge about the sizes and
quasiparticle masses from the ARPES and quantum oscillations data,
we suggest that bulk FeSe
at low temperatures has one hole and two-electron bands,
as represented schematically in Fig.\ref{fig4}.
There are some small discrepancies in the absolute values of the cross section areas
extracted from ARPES and quantum oscillations, but these
could be caused by field-induced Fermi surface effects in quantum oscillations
or possible surface effects in ARPES.
This interpretation is also consistent with the three-band description
of a recent magnetotransport study on FeSe \cite{Watson2015},
in which a very small high mobility carrier was also detected, in agreement with
the size of small electron band from our ARPES data.
It is important to note that our measured  electronic structure
of bulk FeSe is similar to that of multilayers of FeSe on SrTiO$_3$,
in which the splitting of bands at the M point was wrongly assigned to a possible
formation of a spin-density wave \cite{Tan2013}. However, the FeSe monolayer is rather different and its electronic and high $T_c$ superconducting state
was suggested to be influenced by the SrTiO$_3$ substrate \cite{Lee2013}.

One important finding from our the experimental studies
of the electronic structure of FeSe is that the band structure calculations significantly
overestimate the size of the Fermi surface of FeSe, even in the tetragonal phase.
Shrinking of Fermi surfaces in other iron-based superconductors has  been assigned to the interband coupling to a bosonic mode in LaFePO, \cite{Ortenzi2009}, and/or the strength of the antiferromagnetic correlations close to a quantum critical point in BaFe$_2$(As$_{1-x}$P$_x$)$_2$ \cite{Shishido2010}.
It is clear that in FeSe,  the significant shrinking of the Fermi surface, combined
with the strong renormalization effects both in ARPES and quantum oscillations
 as evidenced by the relatively large effective masses of 4-8 $m_e$
 and the orbital-dependent correlations (largest for the $d_{xy}$ band)
suggest that the electronic correlations  significantly affect the
electronic structure of FeSe.

{\bf The nematic susceptibility of FeSe.}
A clear manifestation of a nematic Fermi surface is
its strong in-plane anisotropy in transport properties and sensitivity
to external parameters, in particular in-plane strain.
The resistivity anisotropy is determined
by both the electronic structure and the scattering,
 and the expected Fermi surface deformation
 give rise naturally to anisotropic electronic properties,
 whereas the spin-nematic ordering leads to an anisotropy of the electron scattering \cite{Chu2012}.

Fig.\ref{fig4}k shows the induced change in resistivity to in-plane strain
measured by the $2m_{66}$ parameter from elastoresistance tensor in FeSe,
which provides a direct measure of the electronic nematic order parameter,
as detailed in Fig.SM5 and in Ref.\onlinecite{Kuo2013}. We observe
a large increase in $2m_{66}$ approaching $T_s$, similar to large divergent behaviour
observed in Ba(Fe/Co)$_2$As$_2$ \cite{Kuo2013,Fisher}, but of even larger magnitude.
 The data on FeSe can be well-described
 by a fit to the function $2m_{66}=A/(T-T^*)+A_0$, which gives $T^*=66(1)$~K
 and $A_0$=-31(3) (solid line  in Fig.\ref{fig4}k).
Recently, nematic susceptibility measurements of the elastic shear modulus, which probe the lattice response to strain,
suggest that the structural transition in FeSe is accompanied by a large shear-modulus softening, identical to that of underdoped Ba(Fe,Co)$_2$As$_2$, implying a very similar strength of the electron-lattice coupling,
which is much smaller than the electronic response \cite{Bohmer2014}.

The sign of $m_{66}$  is also opposite to what is found in the electron-doped pnictides, but similar to FeTe \cite{Zhang2012b}, where it was suggested that the resistance along the $a$ (AFM) direction is larger than that along $b$ axis (FM direction)\cite{Zhang2012b}. However,  a small sign-change of the in-plane anisotropy
was also found for highly hole doped (Ba/K)Fe$_2$As$_2$ \cite{Blomberg2012}.
This may suggest that the positive sign of $m_{66}$ may be either a general feature of chalcogenides, as opposed to
most of the pnictides, or it is also possible that the development of the anisotropic properties in all these systems
may be driven by a different mechanism.

Below the structural transition, $T_s$, the behaviour of $m_{66}$ is rather striking, having an almost
 linear temperature dependence with no sign of saturation, as the degree of ellipticity grows larger (see Fig.\ref{fig4}l);
it changes sign around 65~K, which, coincidentally, is the same
 scale as the value of $T^*$ determined earlier. While the interpretation of nematic susceptibility below $T_s$
  may be difficult due to domain formation and the fact that the nematic order parameter now takes a finite value, we suggest that one possible explanation for this crossover may be linked to
  the development of anisotropic scattering from spin-fluctuation, which become strong {\it below} $T_s$ \cite{Baek2014,Bohmer2014}.
 Sign-change of the in-plane anisotropy
has been found between the electron and hole-doped BaFe$_2$As$_2$ \cite{Blomberg2012},
 being assigned to differences in the spin fluctuations
 scattering rates corresponding to different Fermi
velocities at the hot-spots for electron- and hole-doped systems \cite{Breitkreiz2014,Blomberg2012}.
Magnetotransport studies in FeSe also suggest that anisotropic
scattering may develop below $T_s$ \cite{Watson2015}.

As the size of our measured nematic susceptibility
is much larger than the response of the lattice \cite{Bohmer2014}, we suggest
that the structural transition in FeSe is electronically driven and is an instability of the electronic structure which breaks tetragonal symmetry, with the lattice orthorhombicity simply responding to these
  electronic changes. Furthermore, the absence of spin-fluctuations above $T_s$ indicates that the structural transition in FeSe is not magnetically-driven \cite{Baek2014,Bohmer2014}. Thus, our resistivity anisotropy measurements favor
  an orbital/charge ordering scenario, which show a strong splitting of the bands with $d_{xz}$ and $d_{yz}$ orbital character,
  (Fig.\ref{fig2} and summarized in Fig.\ref{fig4}m) and it is likely responsible
  for the in-plane Fermi surface deformation which give rise to significant anisotropy in the in-plane electronic structure.

  While the orbital ordering scenario is a likely contender to explain the existing data on FeSe,
  the Fermi surface deformation and its relatively constant volume conservation (small changes occur
  due to the temperature dependence of the lattice parameters) bears the similarities of a $d$-$wave$
  Pomeranchuck instability. An isotropic Fermi liquid in the presence of sufficiently strong interactions was predicted by Pomeranchuk to be unstable \cite{pomeranchuk_1958a}  and as a signature of this instability the Fermi surface would spontaneously deform, changing its shape or topology to lower its energy and break the rotational symmetry, similar
  to what is observed in FeSe at the Fermi level. The two scenarios are likely to generate different kinds of pairing
  interactions either at finite momentum \cite{Kontani2010} or at zero momentum \cite{Halboth2000}
  and future theoretical work will need to address these issues.

In summary, we report a comprehensive study
of the development of the nematic phase in FeSe.
The Fermi surface of FeSe undergoes a spontaneous distortion
from fourfold-symmetric to two-fold symmetric elliptical pockets.
 The symmetry breaking arises from the electronic degrees of 
 freedom and, in the absence of magnetism,
it is likely to be caused by orbital ordering in the presence of strong interactions.
The elongated Fermi surface causes strongly anisotropic electronic properties
and enhanced nematic suceptibility. This nematic electronic
 phase is that from which superconductivity emerges, which
 is suggested to have a two-fold gap symmetry  \cite{Song2011}.
While interactions favoring orbital
ordering dominate near $T_s$,
 magnetic fluctuations grow towards $T_c$,
 and may still assist the superconducting pairing.
 Moreover, the existence of a relatively flat hole band just
below the Fermi level at the M point
raises the question if whether the presence of van Hove singularities
could also play a role in pairing.
%What exactly triggers the electronic Pomeranchuck instability in bulk FeSe could be linked
%to the strength of electronic correlations and the complex interplay between orbital and lattice effects.
Given the observed small and strongly anisotropic Fermi surfaces with low carrier
densities, it is perhaps not surprising that the physical properties,
including the superconducting and structural transitions
 are susceptible to external parameters (e.g. pressure, strain, doping)
so that controlling these one could turn FeSe into a high-temperature superconductor.

{\bf Acknowledgements}
We acknowledge fruitful discussions with B. Andersen, A. Chubukov, R. Fernandes,
P. Hirschfeld,  L. Gannon, Z. Liu, L. de'Medici, J. C. A. Prentice, R. Valenti, I. Vekter, T. Shibauchi
and P. Dudin. This work was mainly supported by EPSRC (EP/L001772/1, EP/I004475/1, EP/I017836/1).
We thank Diamond Light Source for access to
beamline I05 (proposal number SI9911) that contributed to the results presented here.
Part of the work was performed at the HFML, member of the
European Magnetic Field Laboratory (EMFL).
The authors would like to acknowledge
the use of the University of Oxford Advanced Research Computing (ARC)
 facility in carrying out part of this work.
AIC and AJS are grateful to KITP centre for hospitality
which was supported in part by the National Science Foundation under Grant No. NSF PHY11-25915.
YLC acknowledges the support from the EPSRC (UK) grant EP/K04074X/1 and
a DARPA (US) MESO project (no. N66001-11-1-4105).
AIC acknowledges an EPSRC Career Acceleration Fellowship (EP/I004475/1).

\bibliography{FeSe_29jan15b}

\end{document}